\documentstyle[aps,prl,epsf,floats]{revtex}
\begin{document}

\draft
\wideabs{

\title{Observability of quantum phase fluctuations in 
cuprate superconductors}

\author{Hyok-Jon Kwon$^1$, Alan T. Dorsey$^2$ and P. J. Hirschfeld$^2$ }
\address{$^1$Department of Physics and Center for Superconductivity
Research, University of Maryland, College Park, MD 20742-4111}
\address{$^2$Department of Physics, University of Florida, Gainesville,
FL 32611-8440}
\date{January 4, 2001} 

\maketitle
\begin{abstract}
We study the order parameter phase fluctuation effects in cuprate
superconductors near $T=0$, using a quasi-two-dimensional $d$-wave BCS
model. An effective phason theory is obtained which is used to estimate
the strength of the fluctuations, the fluctuation correction to the
in-plane penetration depth, and the pair-field susceptibility. We find
that while the phase fluctuation effects are difficult to observe in the
renormalization of the superfluid phase stiffness, they may be observed in
a pair tunneling experiment which measures the pair-field susceptibility.
 \end{abstract}
}
Underdoped cuprates exhibit large deviations from the predictions of
BCS mean field theory, including a large gap $\Delta$ which does not scale
with the transition temperature $T_c$, and a ``pseudogap" feature
in the normal state\cite{pseudogap}.   These facts, together with 
the empirical scaling of the 
superfluid phase stiffness (SPS) and $T_c$ with 
doping\cite{Uemura},  have motivated a picture of
the cuprates in which the discrepancies are attributed to 
strong phase fluctuations  \cite{Emery}.  
The estimated phase fluctuation energy scale is smaller than the gap (pairing)
energy scale in these materials, and 
within this picture $T_c$ is determined by the
small SPS rather than the gap energy.  The pseudogap       
is ascribed to a precursor pairing amplitude whose 
phase coherence is destroyed above $T_c$ \cite{ours}. 

As many other explanations for these effects have been put forward,
it is of great interest to devise tests of the phase fluctuation
scenario which distinguish it from others.  
Some  evidence for {\it thermal} phase fluctuation effects 
was provided by Corson {\it et al.} \cite{Corson}, 
who observed unusual conductivity resonances in 
underdoped $\rm Bi_2Sr_2CaCu_2O_{8+\delta}$
(Bi2212) near $T_c$ and analyzed their data
using two-dimensional Kosterlitz-Thouless-Berezinskii
dynamics in the terahertz range. 
It has been claimed that phase fluctuations
are the dominant excitations even at 
low temperatures $T\ll T_c$ \cite{Tlinear}, determining the well-known
linear-$T$ dependence of the penetration depth\cite{Hardy}.
However, Millis {\it et al.}\cite{landau} have shown that {\it quantum} phase 
fluctuations cannot account for this behavior.  
Quantum phase fluctuations do have important consequences for 
the superconductor-insulator  transition\cite{SIT},
and the $c$-axis optical conductivity\cite{Ioffe}.
The above probes provide indirect observation of the phase
fluctuation effects through electronic observables which are modified by the
phase fluctuations.
In this paper, we search for more direct probes of phase fluctuations
within a model of quasi-two-dimensional
$d$-wave BCS superconductors with interlayer Josephson coupling
near optimal doping.

The in-plane SPS can be expressed as
$D_{ab}=n_{s,ab}\hbar^2/ 4md$ where $n_{s,ab}$ is the planar 
superfluid electron density, 
$m$ is the effective mass of the quasiparticle, and $d$ is the
interplanar spacing. Because $D_{ab}$ is determined by the quasiparticle
properties which are not strongly renormalized by the phase fluctuations,
we find that although the renormalization of the Debye-Waller factor $\langle
e^{i\phi}\rangle $ is relatively strong,
both the SPS at $T=0$
and its temperature corrections are weakly
renormalized, in contrast to the case of a Josephson junction array  (JJA)
model\cite{Tlinear,Paramekanti}.
We also consider an experiment to measure the excess current in 
a tunnel junction, which can be directly related to the pair-field 
susceptibility \cite{Carlson} 
$\chi=-i\theta(t)\langle[\Delta({\bf r},t),\Delta^\dagger({\bf 0},0)]\rangle $.
For the cuprates we find that this current is experimentally 
observable, due to the combination of large phase fluctuations and
a low-lying $c$-axis plasmon mode. We predict a pronounced peak in the
in the excess current at the $c$-axis Josephson plasma frequency
$\omega_c$. 

We begin with a continuum BCS model with $d$-wave pairing symmetry in
an isolated two-dimensional layer at temperature $T=1/\beta $ 
in the superconducting state: 
\begin{eqnarray}
& &S_{\rm 2D} = \int_0^\beta d\tau \left\{\int d^2x\sum_\sigma
c^{\dag}_\sigma \left(\partial_\tau -{\nabla^2 \over 2m_0} -\mu\right)
\right. c_\sigma  \nonumber \\
&&\quad +\int d^2R ~d^2r\bigg{[} \Delta ({\bf R,r},\tau)
c^{\dag}_\uparrow({\bf R+r}/2,\tau) c^{\dag}_\downarrow({\bf R-r}/2,\tau) 
\nonumber\\
&&\quad + {\rm h.c.} +{1\over g}|\Delta({\bf R,r},\tau)|^2\bigg{]}\Bigg{\}}, 
\label{Act1}
\end{eqnarray}
with the interaction strength $g>0$. Here and throughout the paper we
set $\hbar =c= k_B =1$ for convenience except when numerical values are
estimated. The gap has the property that 
$\int d^2r ~e^{-i\bf p\cdot r}\Delta({\bf
R,r},\tau) =\Delta({\bf R},\tau)(p_x^2-p_y^2)/p^2$.
Then we factor the  pairing field as
$\Delta ({\bf x},\tau)=|\Delta({\bf x},\tau)|e^{i\phi
 ({\bf x},\tau)}$;
in what follows we will assume that the amplitude of the order parameter 
is constant, $|\Delta({\bf x},\tau)|=\Delta $, and focus on the 
phase degree of freedom, $\phi $.
In order to decouple the 
$\phi  $ field from the order parameter amplitude, we perform a 
singular gauge transformation
$ \psi_\sigma({\bf x},\tau) = c_\sigma({\bf x},\tau)  
e^{-i\phi ({\bf x},\tau)/2}$, with $\psi_\sigma$ the field operators for
the transformed quasiparticle. 
The phase-quasiparticle coupling terms are then 
\begin{equation}
S_I = \int d^2x~\int_0^\beta d\tau \left[{\nabla \phi \over 2}\cdot
\hat{\psi}^\dag {i \over m}\nabla \hat{\psi} 
 + {(\nabla \phi)^2\over 8m}\hat{\psi}^\dag \hat{\tau}_3\hat{\psi}
\right]
\label{couple}
\end{equation}
where $\hat{\psi}$ is the Nambu spinor.
The wavevector of the phase  fluctuations has an
upper cutoff $\Lambda_c$ of order 
$ \xi_0^{-1}\sim\Delta /v_F$ since beyond 
this momentum scale the mean-field assumption breaks down.
Here we consider only the effect of longitudinal phase fluctuations
since the production of vortex pairs is energetically unfavorable near
$T=0$ and far away from the insulator transition.
In order to study a realistic model, we consider such layers of 
two-dimensional superconductors with an interlayer distance $d$ and a
weak Josephson tunneling ($J$)  between adjacent layers, and 
a three-dimensional Coulomb interaction 
$V({\bf q})= 4\pi e^2/\epsilon_b q^2$,
where ${\bf q} = ({\bf q}_\parallel,q_\perp)$, with ${\bf q}_\parallel$  
and $q_\perp$ the in-plane and $c$-axis components of ${\bf q}$, 
and $\epsilon_b$ the background dielectric constant.  
After integrating out the fermions we can obtain the effective 
phase-only action; the Gaussian term is
\begin{eqnarray}
S^{(2)}[\phi]&=& T\sum_{\omega_n,{\bf q}}{ \omega_n^2+\omega_p^2({\bf
q})\over 8V({\bf q})}\phi ({\bf q},\omega_n) \phi (-{\bf q},-\omega_n)
\label{second}
\end{eqnarray}
with $\omega_n=2\pi n T$ the 
bosonic Matsubara frequencies. 
The plasma frequency $\omega_p$ is defined through
$\omega_p^2({\bf q})=(\omega_{ab}^2 q_\parallel^2+\omega_c^2 q_\perp^2)
/(q_\parallel^2+q_\perp^2)$
where $\omega_{ab} = \sqrt{4\pi n_{ab}e^2/\epsilon_b md}$, 
$n_{ab}$ is the  planar charge-carrier density at the plasma resonance
frequency, and 
$\omega_{c}=\sqrt{4\pi Jd^2e^2/\epsilon_b}$ is 
the $c$-axis Josephson plasma
energy.  Equation (\ref{second}) gives the correct plasma spectrum 
for a layered superconductor, with 
$\omega _{ab}$ the planar plasma frequency at $T=0$ which is 
0.44--1.4~eV\cite{wp},  and $\omega_{c}$ is the $c$-axis plasma 
frequency which is about 0.6~meV in Bi2212\cite{JPR} 
(larger in other materials), 
with an interlayer distance $d \approx 1.5~ {\rm nm}$ and 
a dielectric constant of $\epsilon_b \approx 10$\cite{JPR}.  
The in-plane SPS $D_{ab}= n_{s,ab}/4md$
 can be read off from Eq.~(\ref{second})  as the coefficient of 
$(\nabla_{ab}\phi )^2/2$ in the $\omega \rightarrow 0$ and 
$|{\bf q}|\rightarrow 0$ limit.
The quasiparticle damping term, which will broaden the
plasma mode in Eq.~(\ref{second}), has been omitted, but the effect is
known to be small even in the case of the $d$-wave quasiparticle
spectrum \cite{JPR,Hwang},  and it will not  affect the order of magnitude
estimate of the correction to the SPS  in what follows.

We first estimate the strength of the phase fluctuation from 
Eq.~(\ref{second}). One measure of the strength is the Debye-Waller factor
$\alpha = e^{-\langle \phi ^2 \rangle}$, which for instance can be
determined from the $c$-axis optical conductivity \cite{Ioffe}. 
For our model, at $T=0$ we have  
\begin{equation}
\langle \phi^2 \rangle = \sum_{\bf q} 
{2 V({\bf q})\over \hbar\omega_p({\bf q})} 
\approx {2\Lambda_ce^2/\epsilon_b \over  \hbar\omega_{ab}}. 
\label{phi} 
\end{equation} 
From this expression we see that the size of the quantum phase 
fluctuations is determined by the ratio of the Coulomb energy of a
Cooper pair to the plasma energy; in the cuprates the short 
coherence lengths and small superfluid densities conspire to 
enhance these fluctuations.  
For instance, 
assuming  $\xi_0\approx 20~{\rm \AA}$ and with the parameters given above, 
we estimate from Eq.~(\ref{phi}) that 
$\langle \phi^2 \rangle$ ranges from $0.1$ to $1$,
which is a sizable number compared to, for instance, $10^{-3}$ in Pb.
In this paper we study the effect of these strong
phase fluctuations in the BCS model given in Eq.~(\ref{Act1}), 
which is more appropriate near optimal doping, far away from the 
insulator transition. 

\begin{figure}[t]
\epsfxsize=8.0cm 
\begin{center}
\centerline{\epsfbox{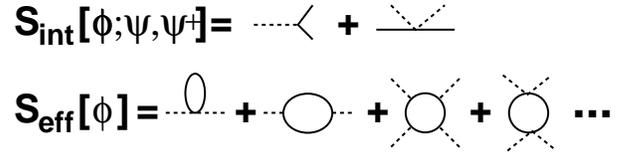}}
\end{center}
\vspace{-0.5cm}
\caption{The schematic diagram of the quasiparticle-phase field
coupling
and the effective action of the phase. The solid lines represent the
quasiparticle and the dashed lines the phase field.} 
\vspace{-0.4cm}
\label{fig:vertex}
\end{figure}

\begin{figure}[t]
\epsfxsize=8.0cm 
\begin{center}
\centerline{\epsfbox{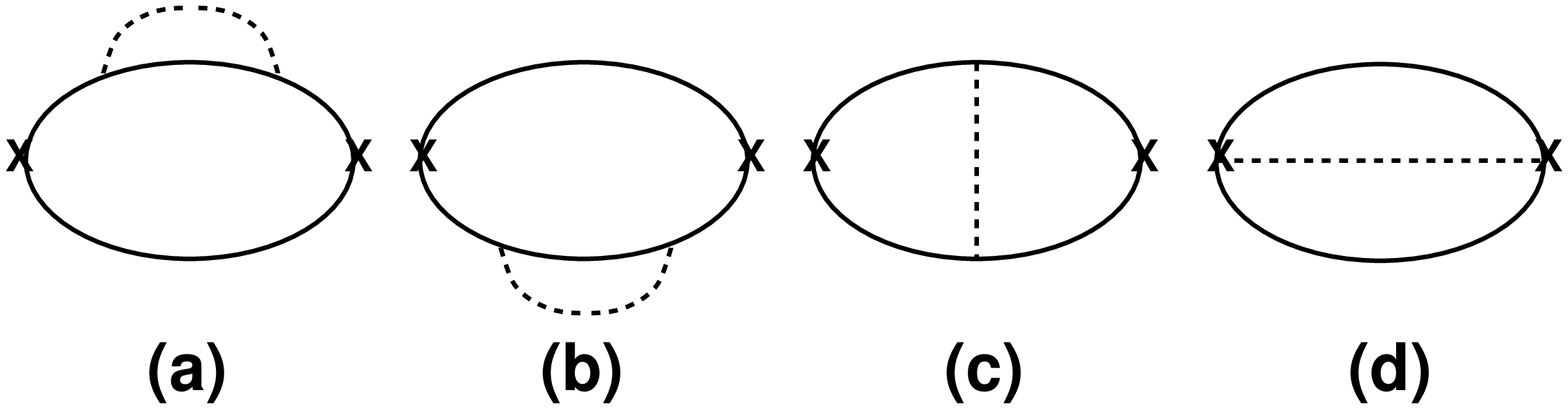}}
\end{center}
\vspace{-0.5cm}
\caption{Correction to the superfluid phase stiffness in one loop
expansion in the phase field.} 
\vspace{-0.4cm}
\label{fig:loop}
\end{figure}

Since we are interested in the renormalization of the SPS, 
we need to go beyond the quadratic expansion in
Eq.~(\ref{second}). For instance, in the JJA model,
the effective SPS can be substantially renormalized due to
the non-trivial potential of the form $\cos(\phi_i - \phi_j)$ 
\cite{Paramekanti}. 
In our model, higher-order terms can be determined by expanding
Eq.~(\ref{couple}) and integrating out the fermions with a $d$-wave
gap. Each $n$-point
vertex of the $\nabla \phi$-field is a fermion loop (see
Fig.~\ref{fig:vertex}). 
Therefore, the renormalization of the SPS and its
temperature dependence is determined by the magnitudes of the fermion
loops. From the new effective theory of the phase fields  thus
obtained, 
\begin{eqnarray}
S_{\rm eff}[\phi]&=& \sum_n
\int d^4x_1...\int
d^4x_{2n}~\Gamma^{(2n)}(x_1,x_2,...,x_{2n}) \label{Seff} \\
&&\times
\nabla_{ 1}\phi(x_1)
\nabla_{ 2}\phi(x_2)...\nabla_{2n}\phi(x_{2n}),\nonumber
\end{eqnarray}
where $\Gamma^{(2n)}$ are the 2$n$th-order phason vertices,
we can estimate the renormalization of the in-plane SPS
by using the one loop expansion in $\phi$ as in Fig.~\ref{fig:loop}.
At $T=0$, we can show that the diagrams in Fig.~\ref{fig:loop} 
(a), (b) and (c) cancel one another in the limit of zero 
external momentum and frequency by using the identity 
$i\partial_\omega \hat{\cal G}({\bf p},\omega )=
\hat{\cal G}^2({\bf p},\omega )$ and
performing the integration by parts in $\omega$. 
Consequently, the only contribution comes from (d).
The correction to the SPS is found to be
\begin{eqnarray}
{\delta D_{ab}\over D_{ab}}  \approx
{e^2\Lambda_c^3\over 6~n_{s,ab}\epsilon_b\hbar\omega_{ab}}.
\end{eqnarray}
Assuming that the in-plane penetration depth is
$\lambda_{ab}=\sqrt{mc^2d/4\pi n_{s,ab}e^2}
\approx 2000~{\rm \AA}$, we estimate that
$\delta D_{ab}/D_{ab}\le 10^{-1}$. 
This should be compared to a 40\% reduction obtained using
the JJA model \cite{Paramekanti}. 
In our model, the  phason vertices are determined 
by the $d$-wave quasiparticle fermion
loops, which are smaller than the vertices of  a JJA model; consequently 
the  renormalization of the SPS is smaller. 

Next we study the temperature dependence of the SPS. 
The BCS theory gives a linear temperature dependence
due to thermal excitations of  quasiparticles,
$D_{ab}(T)\approx D_{ab}(0)-a T $
where $a = v_F\ln(2)/4\pi v_{\Delta}d$. Here $v_{\Delta}$ is the
slope of the gap at the node in momentum space.
We find that the one-loop correction to $a$ is 
${\delta a / a} \approx 
{ e^2v_F\Lambda_c^2\over \pi  \epsilon_b\hbar \omega_{ab}^2}
\sim 10^{-2}$.
The increase in the slope $a$  due to phase fluctuations is therefore 
hardly a measurable quantity, in agreement with the 
JJA result \cite{Paramekanti}.
Additional temperature dependence can
be obtained from considering  classical phase fluctuations or by 
coupling the phason to a heat bath\cite{Tlinear}. However, classical phase
fluctuation effects are not relevant to our model and the coupling to
the  heat bath leads to only a sub-leading $T^2$ correction to
the SPS\cite{landau}.

\begin{figure}[t]
\epsfxsize=8.0cm
\begin{center}
\centerline{\epsfbox{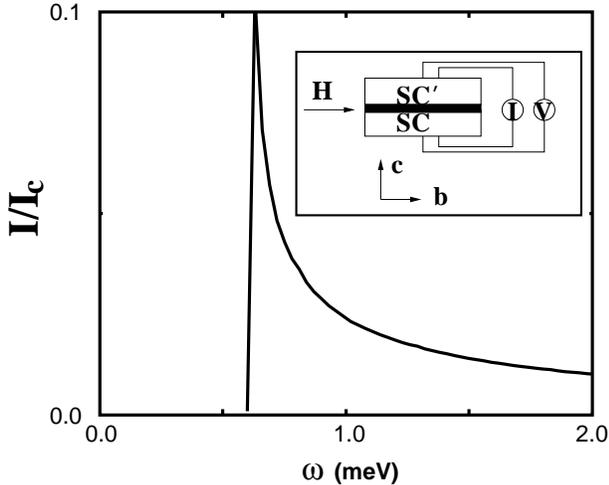}}
\vspace{-0.5cm}
\end{center}
\caption{Excess pair tunneling current where $\omega_c =0.6~\rm meV$
and $\alpha = 0.75$.
The inset figure illustrates a pair tunneling experiment with two cuprate
superconducting electrodes SC and SC$^\prime$.
 We assume a junction contact
area of $10^{-8}~\rm m^2$, 
 a normal state junction resistance $R_N \approx 30 ~\Omega $,
and an in-plane magnetic field of $H\sim 0.3$ Gauss. [See Eq.~(\ref{Iex}).] }
\vspace{-0.4cm}
\label{fig:tunnel}
\label{fig:current}
\end{figure}

Unlike quasiparticle properties represented by the SPS
as discussed above, the phase fluctuation effects on Cooper
pair properties can be significant. Here we propose an experiment
which can measure the strength and the form of the phase fluctuations
via the pair-field susceptibility.
We consider a $c$-axis tunnel junction between two cuprate superconductors
as illustrated in Fig.~\ref{fig:tunnel}\cite{Janko}. 
The Josephson coupling between the 
phases (denoted by $\phi $ and $\phi ^\prime$) of the two superconductors will
lead to the usual Josephson current oscillating at a frequency of
$2eV/\hbar$ if there is a potential difference $V$ across the junction. There
will also be a quasiparticle tunneling current. In addition,
an {\it excess} current will flow due to the Josephson coupling of the 
superconducting pair-field of one superconducting electrode to the 
{\it fluctuating} pair-field of the other. 
To isolate the excess current a small magnetic field is applied
parallel to the junction to suppress the Josephson current, 
and the quasiparticle
tunneling current must be modeled and subtracted \cite{Carlson}. 
The excess current is interesting because it
can be related to the pair-field susceptibility at a frequency
$2eV/\hbar$ \cite{Scal}, and can thus provide information about the 
spectrum of phase fluctuations. 
 
To specialize the experiment to our case, we suppose that both of the
electrodes are identical with a gap $\Delta$ and ignore the
fluctuations in the amplitude of the order parameter assuming that $2eV
\ll \Delta$. For simplicity, we consider a junction in the $ab$ plane
(at $z=0$) of dimensions $L_x\times L_y$,
with a magnetic field $H_y$ in the $b$-direction, and we will
work at zero temperature. If we assume
that the thickness of the electrodes is larger than $\lambda_c$,
the $c$-axis penetration depth, the Josephson coupling Hamiltonian 
for a phase difference 
$\delta\phi({\bf r},t) \equiv \phi({\bf r},t) - \phi'({\bf r},t)$ is
\begin{eqnarray}
H_J &=& {E_J\over 2 S}e^{-i\omega t}\int d^3 r~e^{iq_x x}
 e^{ i\delta\phi({\bf r},t)} \delta(z) +{\rm h.c.},
\end{eqnarray}
where $E_J$ is the Josephson coupling energy, $S=L_xL_y$ is the junction
contact area,  
$q_x \approx 4eH_y\lambda_c /\hbar c$ \cite{Scal}, and $\omega = 2eV/\hbar$. 
The current through the junction is
\begin{equation}
I = - {2e E_J\over \hbar S} {\rm Im}\ 
e^{-i\omega t} \int d^3 r~e^{iq_x x}
\langle e^{i\delta\phi({\bf r},t)} \rangle \delta(z) .
\end{equation}
If we calculate the current to zeroth order in $H_J$, and carry out
the averages with respect to the Gaussian action in Eq.~(\ref{second}), 
we obtain the Josephson current with a critical current 
$\tilde{I}_c = (2e E_J/\hbar)\alpha$
which is renormalized by phase fluctuations through the Debye-Waller
factor $\alpha$ (there is no quasiparticle current in our model). 
By calculating the current to first order in $H_J$ (linear response),
we obtain the excess current, 
\begin{equation}
I_{\rm ex}(\omega,q_x) =
{eE^2_J \over S\hbar^2 }{\rm Im}~D^R(q_x,q_y=0,\omega;z=0),
\end{equation}   
where the retarded pair field susceptibility is 
\begin{equation}
D^R({\bf r},t)=-i\theta(t)\left\langle \left[e^{i\delta\phi({\bf r},t)},
e^{-i\delta\phi({\bf 0},0)}\right]\right\rangle .
\label{corr}
\end{equation}
For $\omega>0$ we have 
${\rm Im}~D^R({\bf q},\omega) = {\rm Im}~D({\bf q},\omega)$,  where 
$D({\bf q},\omega)$ is the Fourier transform of the 
time-ordered correlation function, which for a Gaussian action is
\begin{eqnarray}
D({\bf r},t)&=&-i\alpha^2 
e^{2 \langle T [\phi({\bf r},t)\phi({\bf 0},0)]\rangle}\nonumber\\
& \approx  & -i\alpha^2 \left\{ 1 
+ 2 \langle T [\phi({\bf r},t)\phi({\bf 0},0)]\rangle\right\},
\label{D}
\end{eqnarray}
where the factor of two comes from the two sides of the junction 
and $T$ is the time-ordering operator.
Equation (\ref{D}) assumes an expansion in the small parameter $\ln
\alpha $. 
If we neglect the boundary effects near the junction, we can 
obtain the propagator for the phase fields from the
action in Eq.~(\ref{second}); after an 
analytic continuation, we have
\begin{equation}
\langle \phi({\bf q},\omega) \phi(-{\bf q},-\omega)\rangle
 =  {-4i V({\bf q})\over
\omega^2 - \omega_{p}^2({\bf q}) +i0^{+}}. 
\label{prop}
\end{equation} 

For an {\it isotropic} plasma frequency $\omega_p$ it can be shown that
the excess current consists of a series of
$\delta$-functions at integer multiples of $\omega_p$.  
This result is somewhat academic since in the known 
isotropic superconductors the plasma energy is much larger 
than the gap energy, and we would expect 
amplitude fluctuations and quasiparticle damping to completely
obliterate this effect. 
For an {\it anisotropic} plasma frequency these resonances become
broadened---since the junction is localized at $z=0$, we must
integrate over $q_\perp$, which results in a plasma frequency that 
ranges from
$\omega_{c}$ (when $q_x/q_\perp \rightarrow 0$) to 
$\omega_{ab}$ (when $q_\perp/q_x \rightarrow 0$).  
The result of the calculation is
\begin{eqnarray}
I_{\rm ex} (\omega,q_x) &\approx& \left({2\pi e\tilde{I}_c^2 \over q_x
\epsilon_b L_yL_x\hbar}\right)
 { \theta(\omega - \omega_c)\theta(\omega_{ab} - \omega)
\over \sqrt{ (\omega^2 - \omega_c^2)(\omega_{ab}^2 - \omega^2)}}.
\label{Iex}
\end{eqnarray}
To quench the background Josephson current, a
field may be chosen such that $q_x=2\pi/L_x$;
a result is shown in Fig.~\ref{fig:current}. 

The excess current exhibits a sharp onset at 
$\omega=\omega_c$ with a
peak of the form $(\omega^2 - \omega_c^2)^{-1/2}$. 
This peak would be rounded by
the small  quasiparticle damping which we have not considered here. 
This experiment would serve as a
direct observation of the phase fluctuations and as an alternative
way to measure the $c$-axis Josephson plasma energy\cite{JPR} 
lying below the maximum gap.  In addition, it would
provide a measure of the strength of the phase fluctuations. 
The gapless collective modes studied in Refs. \cite{Carlson,Scal} are
due to the order parameter fluctuations which are rendered visible
near the transition temperature, which are in principle observable in
any superconductors, whereas the quantum phason modes that
we studied are observable as a result of the intrinsically strong
phase fluctuations and the quasi-two-dimensionality of cuprate
materials. Therefore, the result shown in Fig. \ref{fig:tunnel} is a
special zero-temperature property of cuprate superconductors. 

We have shown that in the case of quasi-two-dimensional $d$-wave BCS
model that we used here, the resulting  correction to the absolute 
value and the temperature-dependence of the in-plane SPS is minute 
despite the strong phase fluctuations. 
Since  quasiparticle interaction effects can obscure the fluctuation
corrections to the  SPS, it may be difficult to
observe the effects of phase fluctuations on the penetration depth.
As a more direct measurement, we have proposed a pair tunneling experiment 
which can probe the strength
and spectrum of the quantum phase fluctuations. 
We expect that the pair-field susceptibility will show a
pronounced peak at $\omega = \omega_c$.
It will be also interesting to explore the role of the order parameter
fluctuations at the superconductor-insulator transition in the
underdoped regime  via the suggested experiment. 
In the underdoped regime, the simple BCS model fails,
especially at the superconductor-insulator transition which is one
extreme example of the renormalization of the SPS, and the physics of
doped Mott insulators needs to be taken into account. 

We gratefully acknowledge discussions with A. J. Millis and E. H. Hwang.
This work was supported by the NHMFL, NSF DMR-9978547 (ATD),   
NSF DMR-9974396 (PJH), NSF DMR-9815094, and the Packard Foundation (HJK).
 
\vspace{-1.0\baselineskip}

\end{document}